\documentclass[pdflatex,sn-mathphys-num]{sn-jnl}

\usepackage{graphicx}%
\usepackage{multirow}%
\usepackage{amsmath,amssymb,amsfonts}%
\usepackage{amsthm}%
\usepackage{mathrsfs}%
\usepackage[title]{appendix}%
\usepackage{xcolor}%
\usepackage{textcomp}%
\usepackage{manyfoot}%
\usepackage{booktabs}%
\usepackage{algorithm}%
\usepackage{algorithmicx}%
\usepackage{algpseudocode}%
\usepackage{listings}%
\usepackage{lineno}

\theoremstyle{thmstyleone}%
\theoremstyle{thmstyletwo}%

\theoremstyle{thmstylethree}%

\begin{document}
\title[Article Title]{Spatially Parallel All-optical Neural Networks}


\author*[1]{\fnm{Jianwei} \sur{Qin}}\email{qinjw3@sjtu.edu.cn}
\equalcont{These authors contributed equally to this work.}

\author[2]{\fnm{Yanbing} \sur{Liu}}
\equalcont{These authors contributed equally to this work.}

\author[1]{\fnm{Yan} \sur{Liu}}

\author[3]{\fnm{Xun} \sur{Liu}}

\author*[3]{\fnm{Wei} \sur{Li}}\email{wei$\_$li$\_$bj@163.com}

\author*[4, 1]{\fnm{Fangwei} \sur{Ye}}\email{fangweiye@sjtu.edu.cn}

\affil[1]{\orgdiv{School of Physics and Astronomy}, \orgname{Shanghai Jiao Tong University}, \city{Shanghai}, \postcode{200240}, \country{China}}

\affil[2]{\orgdiv{School of Electronic Engineering}, \orgname{Beijing University of Posts and Telecommunications}, \city{Beijing}, \postcode{100876}, \country{China}}

\affil[3]{\orgdiv{Beijing Institute of Space Mechanics and Electricity}, \orgname{China Academy of Space Technology},
\city{Beijing}, \postcode{100094},\country{China}}

\affil[4]{\orgdiv{School of Physics}, \orgname{Chengdu University of Technology},
\city{Chengdu}, 
\postcode{610059},
\country{China}}


\abstract{All-optical neural networks (AONNs) have emerged as a promising paradigm for ultrafast and energy-efficient computation. These networks typically consist of multiple serially connected layers between input and output layers—a configuration we term spatially series AONNs, with deep neural networks (DNNs) being the most prominent examples. However, such series architectures suffer from progressive signal degradation during information propagation and critically require additional nonlinearity designs to model complex relationships effectively. Here we propose a spatially parallel architecture for all-optical neural networks (SP-AONNs). Unlike series architecture that sequentially processes information through consecutively connected optical layers, SP-AONNs divide the input signal into identical copies fed simultaneously into separate optical layers. Through coherent interference between these parallel linear sub-networks, SP-AONNs inherently enable nonlinear computation without relying on active nonlinear components or iterative updates. We implemented a modular 4F optical system for SP-AONNs and evaluated its performance across multiple image classification benchmarks. Experimental results demonstrate that increasing the number of parallel sub-networks consistently enhances accuracy, improves noise robustness, and expands model expressivity. Our findings highlight spatial parallelism as a practical and scalable strategy for advancing the capabilities of optical neural computing.}




\maketitle

\section{Introduction}
All-optical neural networks (AONNs) represent a cutting-edge computational architecture that employs optical signals for end-to-end data processing, utilizing coherent light diffraction to perform spatial light computing and inference\cite{lin2018all,kues2017chip, liu2022programmable, zhou2021large, kulce2021all}. In contrast to conventional electronic neural networks, AONNs achieve orders-of-magnitude faster processing speeds by exploiting the inherent advantages of optical signals\cite{nips1,nips2,nips3,nips4}, including low transmission loss and high bandwidth. These characteristics make them well-suited for high-throughput, low-latency inference tasks\cite{shastri2021photonics,wetzstein2020inference,feldmann2019all}. 
In electronic neural networks, nonlinear activation \cite{nonline1,nonline2,nonline3,nonline4}functions play a crucial role by enabling the model to learn abstract and discriminative feature representations through nonlinear transformations\cite{lecun2015deep,wang2023image}. This nonlinearity enhances the network’s ability to capture complex patterns in the input data and significantly improves feature separability, leading to higher accuracy in classification tasks \cite{krizhevsky2012imagenet,goodfellow2016deep}. 

However, the development of practical AONNs faces a critical challenge: the implementation of efficient all-optical nonlinear activation functions that can match the performance of their electronic counterparts while maintaining the energy efficiency and parallelism inherent to optical systems\cite{hughes2018training, shen2017deep}. Current approaches struggle to achieve both low power consumption and high nonlinearity simultaneously, often requiring either bulky atomic systems \cite{zuo2019all,zuo2021scalability} or high-intensity optical fields that introduce excessive energy dissipation. Moreover, most existing solutions lack the programmability needed to support diverse activation functions (e.g., ReLU, sigmoid) required for modern neural network architectures, and their integration with scalable photonic platforms remains a significant hurdle for large-scale deployment \cite{multi1,multi2,multi3,multi4,multi5} . These limitations collectively constrain AONNs from realizing their full potential in complex machine learning tasks.


The fundamental challenge in advancing AONNs lies in enabling multi-layer optical coupling and increasing the network’s tunable degrees\cite{optnon1,optnon2,optnon3} of freedom (or neuron count) without depending on conventional nonlinear optical components. Thus, recent research has explored several strategies to realize nonlinear computation through linear optical processes\cite{optline2,opyline1}. One such approach employs multi-scattering cavities\cite{xia2024nonlinear}, where optical scattering generates input-dependent speckle patterns that passively mimic nonlinear transformations. However, this method suffers from uncontrollable noise and significant optical attenuation, necessitating high-power laser excitation. Another strategy involves dynamic diffractive layers\cite{yildirim2024nonlinear}, where iterative phase modulation enables superposition-based nonlinear approximations. While this reduces laser power requirements, it introduces hardware limitations due to the need for continuous parameter updates, leading to potential dynamic errors and speed constraints. Although both methods improve neuron density and computational accuracy, their  respective trade-offs in controllability and system stability remain critical challenges for real-world implementation.


To address the scalability bottlenecks in all-optical neural networks (AONNs), we propose a novel spatially \textit{parallel} architecture that fundamentally redefines information processing paradigms. Existing AONN implementations, including all previously studied variants, are exclusively composed of multiple serially connected layers between input and output – a configuration we term spatially \textit{series} optical neural networks, with deep neural networks (DNNs) representing the most prominent examples [Fig.~\ref{fig1}(a)]. In such series architectures, the information flows through a single-channel that are sequentially connecting optical layers, suffering from progressive signal attenuation and, as aforementioned, requiring explicit nonlinearity implementations. In contrast, our parallel architecture divides the input signal into identical coherent copies that are simultaneously processed through independent, multiple optical layers [Fig.~\ref{fig1}(b)]. These parallel linear sub-networks, with their phase-aligned outputs, are subsequently recombined at the final output stage,  where coherent interference between parallel linear sub-networks naturally generates nonlinear computational capabilities. This parallel processing approach not only preserves signal integrity across layers but also eliminates the need for active nonlinear components or iterative updates, thereby establishing a more robust and scalable framework for optical neural computing. This design supports multi-layer architectures with flexible depth and improved scalability, as the network’s representational capacity increases parabolically with the number of layers. The system employs static, trainable optical parameters that remain fixed post-training, eliminating the need for runtime updates and reducing implementation complexity. Moreover, SP-AONNs maintain robust performance under low-power continuous-wave laser operation, offering a promising path toward scalable, low-latency optical learning systems. The inherent interference mechanism between parallel pathways enables complex feature extraction and pattern recognition tasks that were previously unattainable in conventional series architectures, opening new avenues for developing next-generation optical computing systems.


\section{SP-AONNs framework}

\begin{figure}[htbp]
\centering
\includegraphics[scale=0.22]{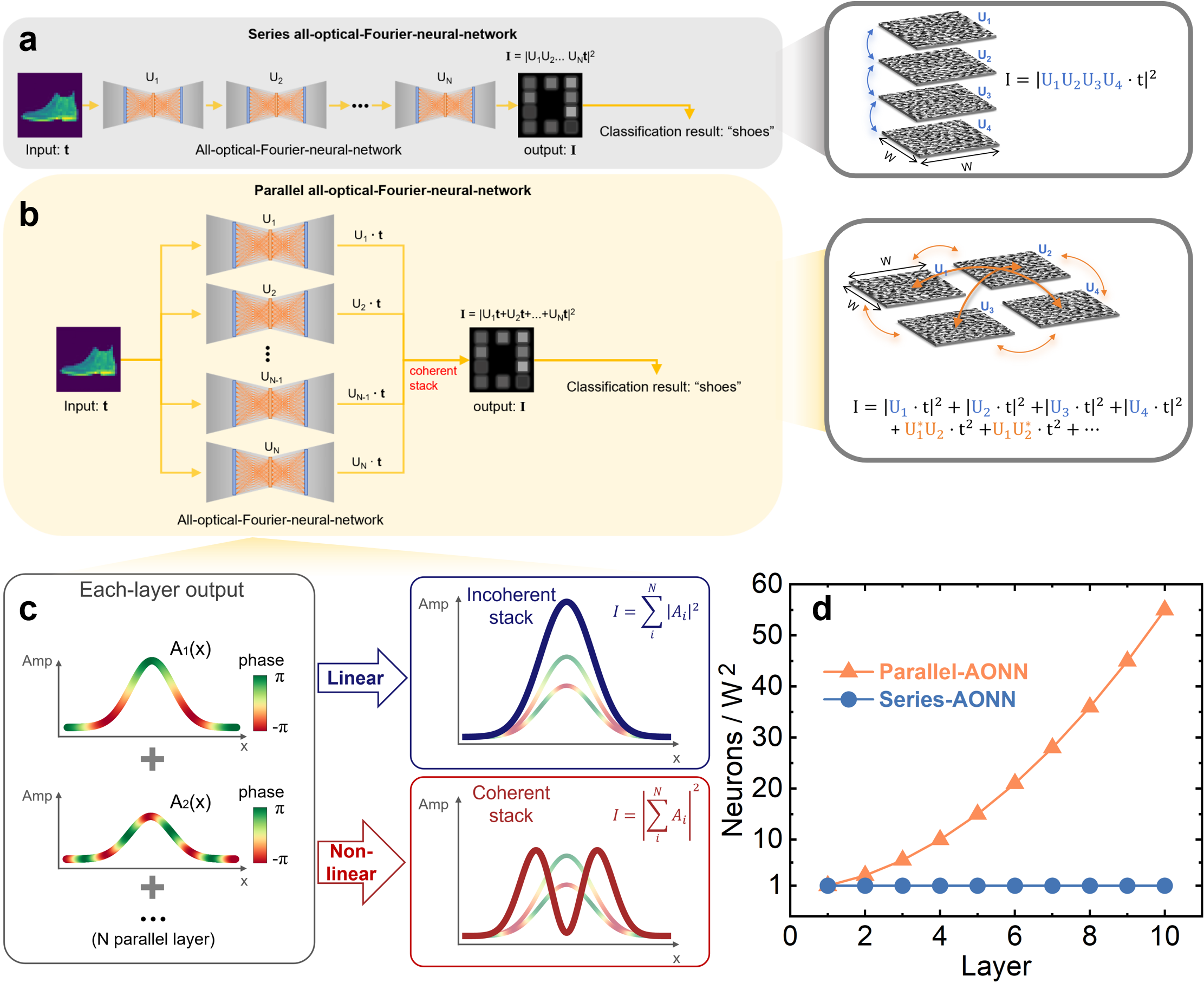}
\caption{(a) Schematic of the serial all-optical Fourier neural network architecture. (b) Schematic of the parallel all-optical Fourier neural network architecture.(c)  Schematic of the coherent-induced nonlinear computational process. (d) Neuron counts in parallel and serial architectures with varying layer numbers.}
\label{fig1}
\end{figure}

A comparison between conventional (series) architectures and our proposed spatially parallel architecture is illustrated in Fig.~\ref{fig1}(a) and Fig.~\ref{fig1}(b). The parallel architecture relies on the coherent superposition of multiple spatially independent optical networks to achieve nonlinear parallelization, thereby creating a more powerful computational architecture. As illustrated in Fig.\ref{fig1}(b), the SP-AONNs employ fan-out of the input image $t$ to generate N identical copies, which are then fed into N parallel single-layer AONN units $\{U_1,...,U_N\}$. Each unit $U_i$ precisely modulates the input image $t$, and its output, $U_i\cdot t$, is eventually coherently stacked through a common-path interferometric setup, yielding the final output light field,
\begin{equation}
    I_{par} = \left| \sum_{i=1}^N U_i \cdot t\right|^2
\end{equation}
The classification results are obtained based on the intensity distribution of the output light field. In comparison, the conventional serial architecture - as illustrated in Fig.~\ref{fig1}(a) - processes the input image $t$ sequentially through N cascaded single-layer AONN ($U_1$ to $U_N$). Therefore, in the absence of an additional nonlinear activation function, the serial network effectively implements a cascade of linear transformations, with the output light field expressed as
\begin{equation}
    I_{ser} = \left|\prod_{i=1}^N U_i \cdot t\right|^2
    \label{eq2}
\end{equation}

In the following, we demonstrate the advantages of parallel architecture by analyzing its neuron capacity enhancement. In optical neural networks, the number of neurons is determined by the independently tunable degrees of freedom within the network\cite{yildirim2024nonlinear}. The input neuron count is defined by the image resolution (e.g., $1000\times1000$ pixels), while the output neuron count depends on the output dimensions (e.g., $150\times 150$ pixels). For the SP-AONNs architecture with N parallel layers, the neuron count in the phase modulation layer comprises both intra-network components, representing independent degrees of freedom within each parallel sub-network that scale linearly with N, and inter-network coupling terms arising from coherent interference between sub-networks that scale quadratically with N (orange curve in Fig.~\ref{fig1}d),
\begin{equation}
    \underbrace{N\times W^2}_{intra-coupling} + \underbrace{C_N^2\times W^2}_{inter-coupling}= (N+\frac{N(N-1)}{2})\times W^2
\end{equation}
(where N represents the number of parallel networks and W denotes the single-network size). In contrast, as explicitly expressed in Eq.~\ref{eq2}, the output light field of conventional serial architectures can always be represented as an equivalent transformation of a single-layer network, maintaining a constant neuron count (blue curve in Fig.~\ref{fig1}d). This fundamental limitation inherently restricts the network's expressive power. Additionally, serial architectures suffer from signal-to-noise ratio degradation caused by exponential light intensity attenuation upon light propagation throughout the networks, which persists even one includes an effective activation function into the structure. These inherent drawbacks make it particularly challenging to construct deep networks using serial configurations,  further highlighting the innovative value of the proposed spatially parallel approach in overcoming network capacity bottlenecks while maintaining light intensity stability.

A central advantage of the SP-AONNs framework in comparison to its series counterpart, lies in the fact that the SP-AONNS features an inherently coherence-induced nonlinear computing mechanism. To appreciate this point, let us consider the scalar diffraction theory\cite{theo1,theo2,theo3}, which tells the output light field $A_i$ of $i$-th sub-network is expressed as:
\begin{equation}
    A_i(x',y') = t(x,y) * h_i(x,y)
\end{equation}
where $*$ denotes the convolution operator, $t(x,y)$ is the input light field determined by the input image, and $h_i(x, y)$ is the point spread function (PSF) defined by the $i$-th AONN channel. In the SP-AONNs, the total output light intensity distribution $I_{\text{total}}$ is given by the coherent superposition of outputs from multiple such channels, expressed as:
\begin{equation}
    \begin{aligned}
    I_{\text{total}}(x', y') = &\left| \sum_{i=1}^N  t(x,y) * h_i(x,y) \right|^2\\
    =&\sum_{i=1}^N \left| t * h_i(x,y)\right|^2 + \sum_{\substack{i,j=1 \\ i \neq j}}^N \text{Re}\left[ (t * h_i(x,y))(t * conj\{h_j(x,y)\} ) \right]
    \end{aligned}
\end{equation}

This expression reveals the nonlinear computational mechanism of the system. As illustrated in Fig.~\ref{fig1}.c, the first term (linear term) represents the independent contributions of each sub-network, offering $N$ degrees of freedom. The second term (interference term) includes $N(N-1)/2$ cross terms, which introduce nonlinear behavior through coherent interference. Physically, the nonlinearity in SP-AONNs arises from two sources:

{\em Spatial coherence-induced nonlinearity--}The interference cross term can be explicitly expanded into its integral form as follows:
\begin{equation}
\label{eq:cross_term}
\begin{aligned}
cross_{ij}(x',y') &\propto \sum_{\substack{i,j=1 \\ i \neq j}}^N\,\text{Re} \Bigg[ \iiiint t(x_1,y_1) t^*(x_2,y_2)\times h_i(x'-x_1,y'-y_1)  \\
&\quad conj\{h_j(x'-x_2,y'-y_2)\}\times d x_1 d y_1 d x_2 d y_2 \Bigg]
\end{aligned}
\end{equation}
This demonstrates that the output depends not only on the local field distribution $t(x,y)$ of the input light field, but also sensitive to the spatial coherence of the input field - specifically, the nonlocal coupling between $t(x_1,y_1)$ and $t(x_2,y_2)$. Such nonlocal coupling is fundamentally unattainable in linear systems and constitutes the origin of nonlinear effects in parallel networks.

{\em Phase-sensitive nonlinearity--} When the phase modulation in the transfer function $h_{i,j}$, $\phi_{i,j}$, varies across sub-networks, the resulting phase differences in $h_{i,j}$ induce interference effects in the cross terms. As shown in Fig.\ref{fig1}.c, this phase-dependent interference mechanism significantly enhances the nonlinear response characteristics.

Therefore, this coherence-dependent nonlinearity enables the system to have effective feature extraction capabilities while relying solely on the coherent superposition of linear optical components, eliminating the need for a nonlinear active function.

\section{Result}

\begin{figure}[htbp]
\centering
\includegraphics[scale=0.25]{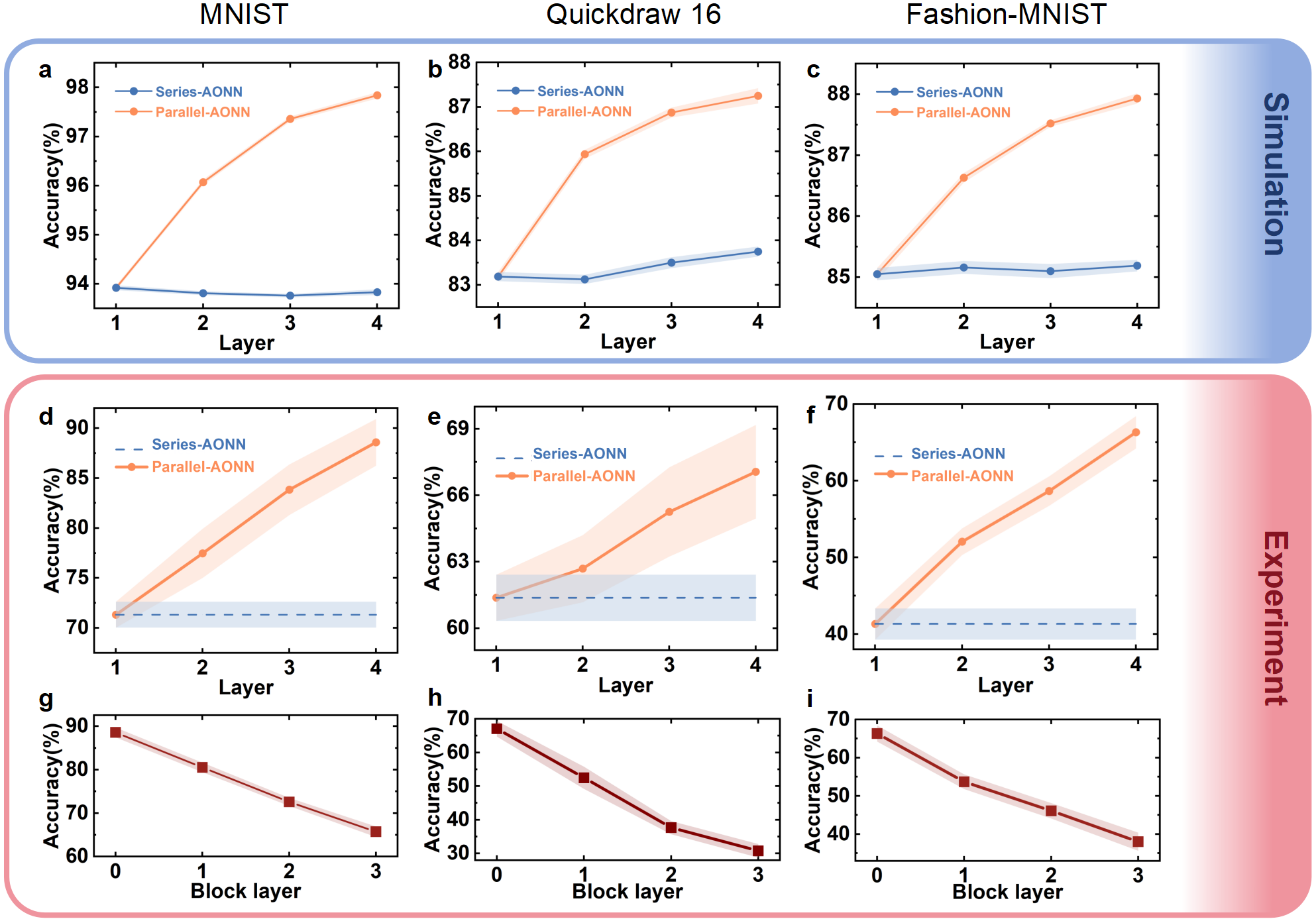}
\caption{(a-c) Simulated accuracy of AONNs with parallel and serial architectures at different layers. (d-f) Experimental accuracy of AONNs with parallel and serial architectures at different layers. (g-i) Experimental classification accuracy when blocking varying numbers of sub-networks in deployed N=4 SP-AONNs}
\label{fig2}
\end{figure}

The preceding theoretical analysis demonstrates that increasing the number of parallel layers 
(i.e., sub-networks) in SP-AONNs effectively enhances the network's degrees of freedom through nonlinear interference effects, which leads to better performance and higher image recognition accuracy. In this section, we conducted a controlled comparison between parallel and serial architectures with identical layer counts, tested on the same image datasets. To be specific, here we implemented SP-AONNS using phase grating beam splitting with multiple 4F optical systems, and the performance was evaluated on three standard benchmark datasets: Digit MNIST, Quickdraw16, and Fashion-Mnist.



The experimental setup and training workflow (Fig.\ref{fig3}b) operate as follows: Input images are loaded into the light field via a DMD. After $8.5~$mm diffraction, a Dammann grating (SLM1) splits the beam into $N$ identical light fields fed into individual sub-networks. Each sub-network contains a Fourier transform layer (1st FL), phase modulation layer (SLM2), and inverse Fourier transform layer (2nd FL). Light fields of sub-networks propagate independently, with outputs coherently superimposed to form the final field. This output is partitioned into 10 (Digit MNIST/Fashion-MNIST) or 16 (Quickdraw 16) classification regions. Predictions derive from power average pooling per region, with labels determined by the argmax operation. Training employs the Adam optimizer, where the loss between predictions and ground truth guides backpropagation to optimize phase modulation layers. 

For clarity, each sub-network is designed to operate independently. To this end, the experimental phase modulation units per subnetwork are set to a size of $W = 200$, additional analysis regarding the effect of sub-network size is detailed in the supplementary material. The light field propagation is modeled using the beam propagation method (BPM). However, directly transferring simulation-trained models to hardware leads to substantial performance degradation due to experimental imperfections. To address this issue, we introduce a lightweight fully connected error-compensation network (10-16-32-16-10) at the output layer. Its output is residually connected to the main network, and joint training is employed to fine-tune the phase modulation layers, thereby mitigating experimental errors. The performance improvements achieved through the compensation network are provided in the supplementary material.

Figure \ref{fig2} shows simulated and experimental classification accuracy for $N=1–4$ SP-AONNs. Consistent with theory, accuracy progressively improves with parallel layers (orange curve), unlike serial architectures (blue curve). SP-AONNs' efficacy is further evidenced by sub-networks' independent classification capability: Channel occlusion tests on a 4-sub-network SP-AONN (Fig.\ref{fig2}g–i) reveal that for MNIST, occluding any single sub-network reduces accuracy from 88.58\% to 80.53\%, while retaining only one operational sub-network (three occluded) maintains 65.72\% accuracy, confirming both standalone discriminative power and architectural robustness.

\section{Discussion}
\begin{figure}[htbp]
\centering
\includegraphics[scale=0.38]{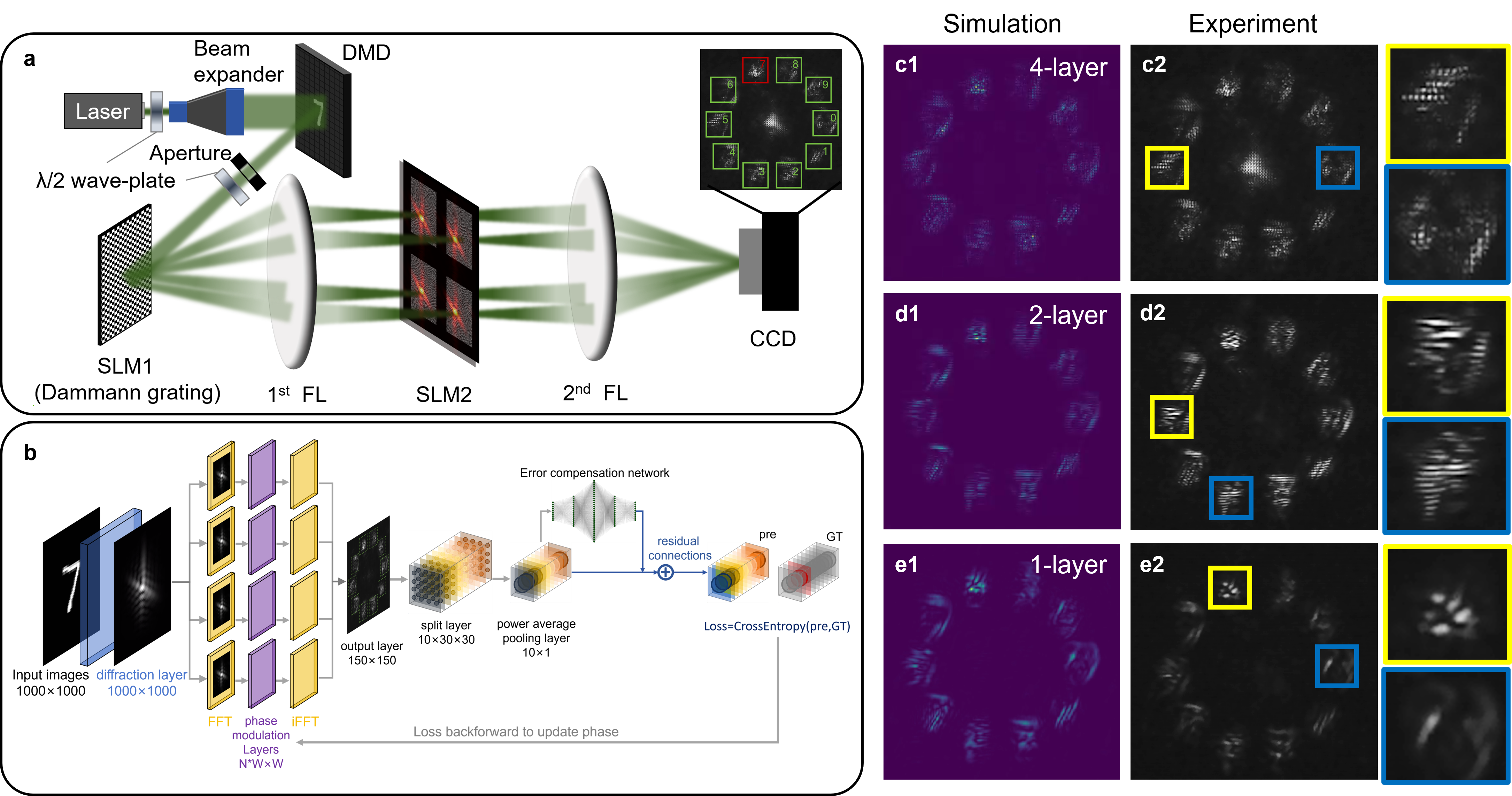}
\caption{(a) Experimental schematic of the SP-AONNs. DMD: digital micromirror device loaded with input images; SLM1: spatial light modulator loaded with Dammann grating map; SLM2: spatial light modulator loaded with a trained phase map; FL: Fourier lens employed as the device for the Fourier transform of the light field. CCD: charge-coupled device employed to capture the output light field. (b) Training schematic of SP-AONNs. (c1-e2) Simulated and experimental output light fields of SP-AONNs, with insets magnifying classification regions to reveal interference-generated lattice patterns.}
\label{fig3}
\end{figure}

The performance enhancement of SP-AONNs stems from the nonlinear computational process of coherent light field superposition, which we verified by examining output field patterns at varying layer counts. For an N=4 SP-AONN with four sub-networks arranged in a $2\times2$ configuration at the Fourier plane, the final output field demonstrates a square lattice pattern (Fig.\ref{fig3}c1–c2) resulting from four-beam interference. An N=2 configuration produces a striped lattice (Fig.\ref{fig3}d1–d2), while N=1 exhibits no lattice structure(Fig.\ref{fig3}e1–e2).

\begin{figure}[htbp]
\centering
\includegraphics[scale=0.19]{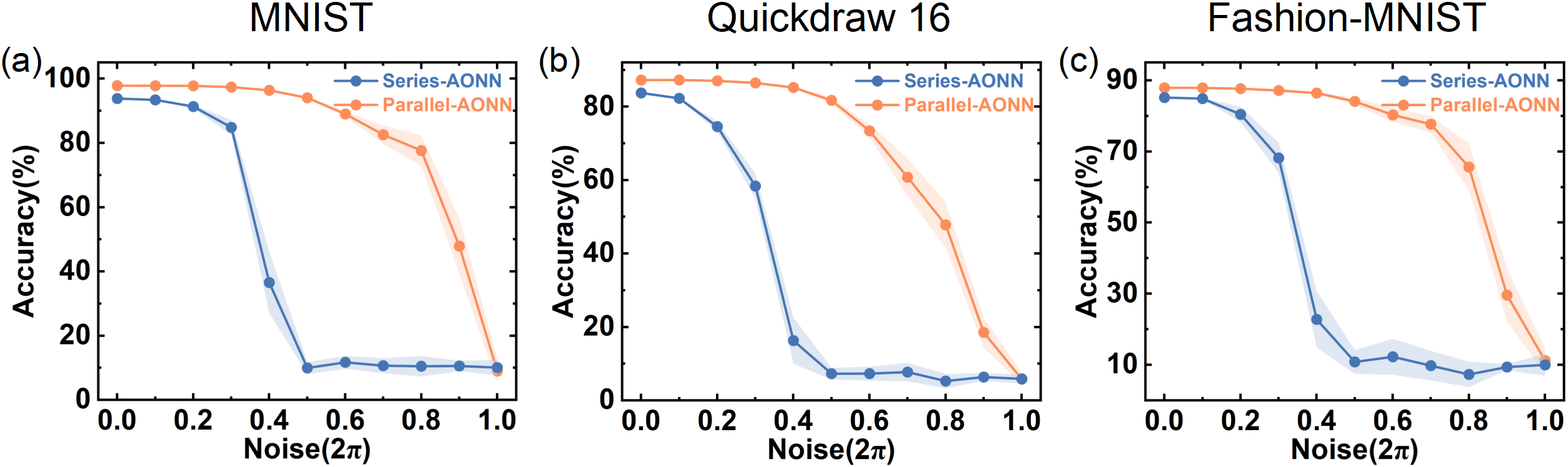}
\caption{Classification accuracy of N=4 parallel and serial AONNs under varying phase noise levels.}
\label{fig4}
\end{figure}

Furthermore, the SP-AONNs framework exhibits superior noise resistance compared to serial architectures due to independent sub-network operation. We evaluated a trained 4-layer sub-network model by introducing phase noise ($0~-~\delta\cdot2\pi$, $\delta\in[0,1]$) to each phase modulation layer, measuring classification accuracy across datasets (Fig.~\ref{fig4}). For Digit MNIST, serial networks degraded to 9.95\% accuracy (complete loss of classification capability) at $1.0\pi$ phase noise, whereas parallel networks maintained 94.07\% accuracy under identical noise conditions. As shown in Fig.\ref{fig4}, classification accuracy remains stable until phase noise intensity exceeds $2\pi$. Since phase noise is a common problem in real-world optical neural networks, the SP-AONNs framework provides an effective way to maintain accuracy. 

\section{Conclusion}
In summary, in contrast to the extensively studied spatially serial optical neural networks (ONNs), we propose a spatially parallel architecture. This design uses the coupling between its parallel sub-networks to generate an inherent, structure-induced nonlinear computational capacity. This intrinsic nonlinearity effectively eliminates the dependence on conventional nonlinear optical materials, offering a highly scalable and hardware-efficient pathway for optical neural computing. Experimental results confirm that increasing the number of parallel sub-networks enhances classification accuracy while also conferring remarkable resilience to phase noise.

Looking ahead, the proposed parallel framework demonstrates significant potential for broad applicability across a diverse range of ONN architectures. This includes not only the optical Fourier neural networks used in this work, but also diffractive ONNs and integrated on-chip photonic designs. By facilitating complex nonlinear computations using purely linear optical elements, this approach may substantially boost network performance while simultaneously minimizing hardware complexity and associated energy costs.

Furthermore, the parallel architecture's non-accumulative error characteristic is a critical advantage, as it can effectively mitigate noise propagation throughout the optical system, thereby significantly improving overall noise tolerance. Future research directions will explore the integration of partially coherent light sources and other coherence-management techniques to further improve this inherent robustness~\cite{qin2025all, dong2024partial, jia2024partially}. For partially coherent light, the coherent components from different parallel channels can still interfere, preserving the nonlinear activation function's effect. By integrating nonlinear activation functions with partially coherent light sources, we anticipate enhanced network robustness and improved target recognition accuracy. These advancements are essential for transitioning laboratory-scale demonstrations into practical, real-world applications where environmental stability is unpredictable. Thus, this work paves the way for the development of more reliable, efficient, and deployable optical computing systems for next-generation artificial intelligence.

\bibliography{output}

\end{document}